\documentclass{iau} 
\usepackage{graphicx, hyperref, natbib, textcomp, wrapfig}
\usepackage{subcaption}

\title[The analysis of VERITAS muon images using convolutional neural networks] 
{The analysis of VERITAS muon images using convolutional neural networks}

\author[Qi Feng and Tony T.Y. Lin \\
for the VERITAS Collaboration]   
{Qi Feng$^{1,\dag}$ and Tony T.Y. Lin$^1$ \\
for the VERITAS Collaboration$^2$ \\
}
\affiliation{$^1$Physics Department, McGill University, Montreal, QC H3A 2T8, Canada\\[\affilskip]
$^2$\url{http://veritas.sao.arizona.edu/} \\[\affilskip]
$\dag$\href{mailto:qi.feng2@mcgill.ca}{qi.feng2@mcgill.ca}}

\pubyear{2016}
\volume{325}  
\jname{IAU Symposium 325 on Astroinformatics}
\editors{Massimo Brescia, George S. Djorgovski, \\ Eric Feigelson, Giuseppe Longo, \& Stefano Cavuoti}
\begin{document}

\maketitle

\begin{abstract}
Imaging atmospheric Cherenkov telescopes (IACTs) are sensitive to rare gamma-ray photons, buried in the background of charged cosmic-ray (CR) particles, the flux of which is several orders of magnitude greater. The ability to separate gamma rays from CR particles is important, as it is directly related to the sensitivity of the instrument. This gamma-ray/CR-particle classification problem in IACT data analysis can be treated with the rapidly-advancing machine learning algorithms, which have the potential to outperform the traditional box-cut methods on image parameters. 
We present preliminary results of a precise classification of a small set of muon events using a convolutional neural networks model with the raw images as input features. We also show the possibility of using the convolutional neural networks model for regression problems, such as the radius and brightness measurement of muon events, which can be used to calibrate the throughput efficiency of IACTs. 

\end{abstract}

\firstsection 
\section{Introduction}
In the past decade, our understanding of the very-high-energy (VHE; 100~GeV $\lesssim E_{\gamma} \lesssim$ 100~TeV) gamma-ray sky has greatly progressed by the use of stereoscopic imaging atmospheric Cherenkov telescopes (IACTs). These IACTs record the images of an extensive air shower induced by an incident VHE gamma-ray photon or a cosmic-ray (CR) particle. The air-shower images are then analyzed to reconstruct the information of the incident photons or the CR particles, the latter of which form a substantial background in VHE gamma-ray astronomy. 
The ability to separate gamma rays from CR particles is important, as it is directly related to the sensitivity of the instrument. 

A few geometric image parameters, e.g. width and length, of an air-shower image are shown to be effective in discriminating gamma rays from CR background \citep[e.g.][]{Hillas85}. These parameters exploit our knowledge of the well-understood air shower physics, specifically the fact that a CR shower typically produces a wider image as it generally carries a larger transverse momentum as a result of hadronic interactions. 
The long-established analysis method is to apply box cuts to the image parameters \citep[e.g.][]{Hillas85}. It is simple and effective, therefore has been the standard analysis method for two decades. 

However, the box-cuts method mentioned above can be augmented by more complicated analysis methods. 
Some of these advanced methods exploit more details in the air showers or their images, e.g. 3-D reconstruction of air showers \citep[e.g.][]{Lemoine-Goumard06}, adding timing structure \citep[e.g.][]{Aliu09}, fitting a 2-D Gaussian to shower images \citep[e.g.][]{Christiansen12}, using a 3-D maximum likelihood analysis \citep[e.g.][]{Cardenzana15}, and matching image templates \citep[e.g.][]{Vincent15}. 
Others use the same image parameters but more sophisticated classification models compared to box cuts. 
An actively pursed approach of such by the current major IACTs is the application of machine learning models using the image parameters as input features, e.g. Bayesian multivariate analysis \citep[e.g.][]{Aharonian91}, artificial neural networks \citep[e.g.][]{Reynolds93}, random forest \citep[e.g.][]{Albert08} and boosted decision trees \citep[e.g.][]{Acero09, Park15, Staszak15, Krause16}. 

On the frontier of computer vision, machine learning algorithms have improved dramatically in the past decade, thanks to the increasing computing power of CPUs and GPUs, as well as large public databases with hand-labelled images (e.g. \href{http://image-net.org}{ImageNet}). Recently, a particularly powerful algorithm, convolutional neural networks (CNN), has gained popularity for its performance \citep[e.g.][]{Krizhevsky12}. 
A CNN model uses images (arrays of pixel values) as input, and usually consists of several convolutional layers followed by pooling (down-sampling) layers, the output of which are then fed into fully-connected layers, and finally to the output. Dropout layers can be used in between the above layers to regularize the CNN model and prevent overfitting. 
The number of parameters in a CNN model can be very large, easily exceeding $10^5$. Therefore, the training process is computationally intensive, calling for the use of GPUs. 

The image parameters in VHE gamma-ray data analysis greatly reduce the dimension of the data using domain knowledge. However, the use of fitted parameters inevitably leads to information loss.
We propose to apply CNN models to raw images of VHE gamma-ray events, exploiting pixel-level information in the data. 
We show that our simple CNN model works well on single-telescope images, being able to correctly classify ring-like muon images at a nearly perfect area-under-the-curve (AUC) score. We also show that it is possible to use the output of convolutional layers as the input to fully-connected regression layers and produce continuous output values (instead of categorical values).


\section{The VERITAS array and the standard data analysis}
VERITAS (the Very Energetic Radiation Imaging Telescope Array System) is an array of
four IACTs located in southern Arizona \citep[30$^\circ$40'N 110$^\circ$57'W, 1.3 km a.s.l.;][]{Holder11}. It is sensitive to gamma rays in the energy range from 85 GeV to $>$30 TeV with an energy resolution of $\sim$15\% (at 1~TeV). 
Each of the four telescopes is equipped with a 12-m diameter Davies-Cotton reflector comprising 355 identical mirror facets, 
and a 499-pixel photomultiplier tube (PMT) camera covering a field of view of 3.5$^\circ$ at an angular resolution (68\% containment) of $\sim$0.1$^{\circ}$ (at 1~TeV). Coincident Cherenkov signals from at least two out of the four telescopes are required to trigger an array-wide read-out of the PMT signals. The array-level trigger occurs at a typical rate of $\sim$400 Hz. 
Most of these triggers come from CR particles. For a comparison, the brightest steady VHE source, the Crab Nebula, is typically observed by VERITAS at a rate of $<$15 gamma rays per minute, much lower than the CR trigger rate of $\sim$400 Hz. 

In this work, we focus on muon events, which are produced in CR showers and observed as rings or partial rings. 
Selected observations are first analyzed using one of the standard VERITAS data analysis packages named \textit{VEGAS} \citep{Cogan08}. 
The \textit{VEGAS} analysis of muon events follows the procedure described below: 
\begin{enumerate}
\item calculate the brightness-weighted average coordinates as the image centroid, and use them as the initial muon-ring center; 
\item calculate the mean ($\bar{r}$) and the variance ($\sigma^2_r$) of the distances between all image pixels and the initial centroid, and use $\bar{r}$ initial muon radius; 
\item move the initial centroid around by a small step, repeat step $(b)$ and check if the variance $\sigma^2_r$ decreases; if so, update the centroid $\bar{r}$ and the variance $\sigma^2_r$; 
\item repeat step $(c)$ to cover a predefined region around the initial centroid, and return the optimal centroid and radius that minimize $\sigma^2_r$; 
\item check if $>$70\% of the pixels fall into the a predefined accepted annulus e.g. $(\bar{r}-1.5 \sigma_r, \bar{r} +1.5 \sigma_r)$; if so, accept this event as a muon event. 
\end{enumerate}

A double-pass method based on the above procedure is used in the analysis, 
and bad pixels are corrected for. 
We use the analysis described above to label signal events for the training/test data, with an additional requirement that $\bar{r} \geqslant 0.5^\circ$. 

\begin{figure}[h]
\begin{center}
 \includegraphics[width=0.8\textwidth]{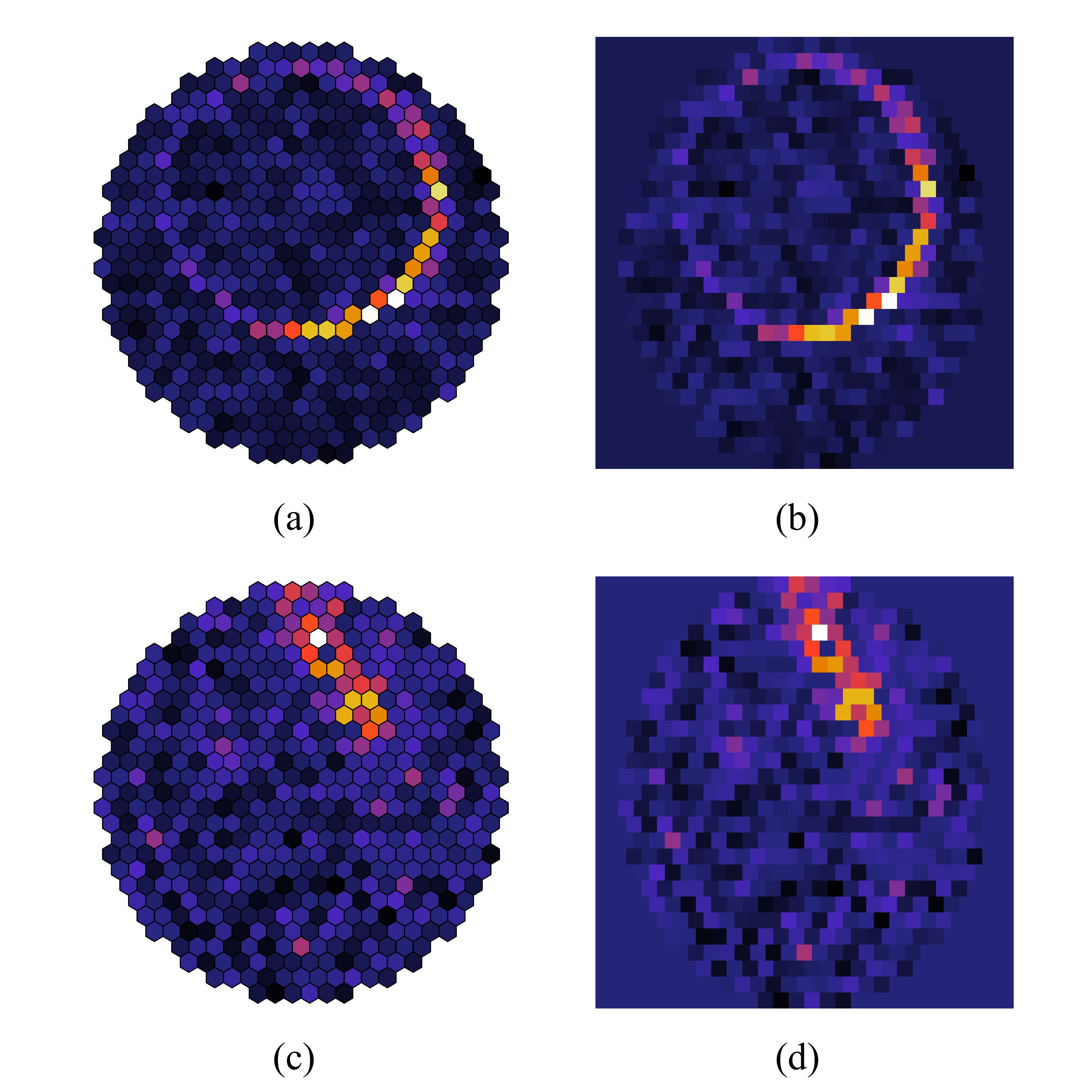} 
 \caption{\small Examples of VERITAS images. The top row shows a muon event, and the bottom show shows a non-muon event. The left column shows the raw images, in which the hexagonal arrangement of the PMTs is apparent, and the right column is the oversampled 54 $\times$ 54 images with a stretch along the vertical direction. }
   \label{fig1}
\end{center}
\end{figure}

Four 30-minute observations are used as training and validation data in this study. Our analysis yields 1597 muon events, which are used as signal events. We randomly chose 4800 events in the remainder of the same observations as background events. 
Two more 30-minute observations independent from the training/validation set are used as test data, which consists of 630 
muon events and 2400 randomly-selected non-muon events. 
The ``true'' signal and background hereafter in this work refer to the \textit{VEGAS}-identified muon and non-muon events. The raw images of two VERITAS events are shown as examples in the left column of Fig.\,\ref{fig1}, where subfigure (a) shows a muon event and subfigure (c) shows a non-muon event. 

\section{The convolutional neural networks classification model, the training, and the evaluation}
%
%
The only preprocessing of the data in the presented model is the image oversampling, which we use to approximately convert the image from its natural hexagonal coordinate (due to the geometry of the PMT arrangement) into a square coordinate. 
We first use a rectangular lattice to cover the pixels, so that each pixel is divided into four equal-sized rectangles. Then we stretch the rectangular lattice to a square one. This converts a 499-pixel hexagonal image into a 54 $\times$ 54 pixel square image, and stretches the image roughly by 15\%. 
In the right column of Fig.\,\ref{fig1}, we show the oversampled images of the same two VERITAS events shown in the left column. 
These images illustrates that the oversampling process only introduces a small amount of stretch, and spatial relation between different pixels is maintained, which is important for CNN models. 
We refer to ``pixel'' as in the oversampled 54 $\times$ 54 space hereafter. 

The oversampled 54 $\times$ 54 pixel images of the signal muon events and the background events are used as the input feature into a convolutional neural networks model, which is implemented using the \textit{keras} Python deep learning library \citep{Chollet13} running on top of \textit{TensorFlow} \citep{Abadi16}. 
The structure of the CNN model is a simplified ``VGG''-style model \citep{Simonyan14}, with only three layers of small filters, average pooling and dropout in between filter layers, and a two-layer fully-connected neural network classification model after the convolutional layers. 

The first convolutional layer has 32 filters, each of which is 6 $\times$ 6 pixel with a 2-pixel stride (roughly corresponding to the physical pixel size 3 $\times$ 3 with stride 1 as a result of the oversampling); 
the second and third convolutional layers are 3 $\times$ 3 with stride 1. The average pooling is performed after each convolutional layer over a 2 $\times$ 2 pixel window, and dropout at a ratio of 25\% is carried out after each pooling layer to regularize the model. The two fully-connected dense layers contain 256 and 64 neurons, respectively. Dropout at a ratio of 50\% is applied after each fully-connected layer. Rectified linear unit is used as the activation function of all the layers (both convolutional layers and fully-connected ones) except the output layer, which uses a sigmoid function as the activation function. 

%
\begin{wrapfigure}{r}{0.5\textwidth}
\begin{center}
\vspace*{-0.8 cm}
 \includegraphics[width=0.5\textwidth]{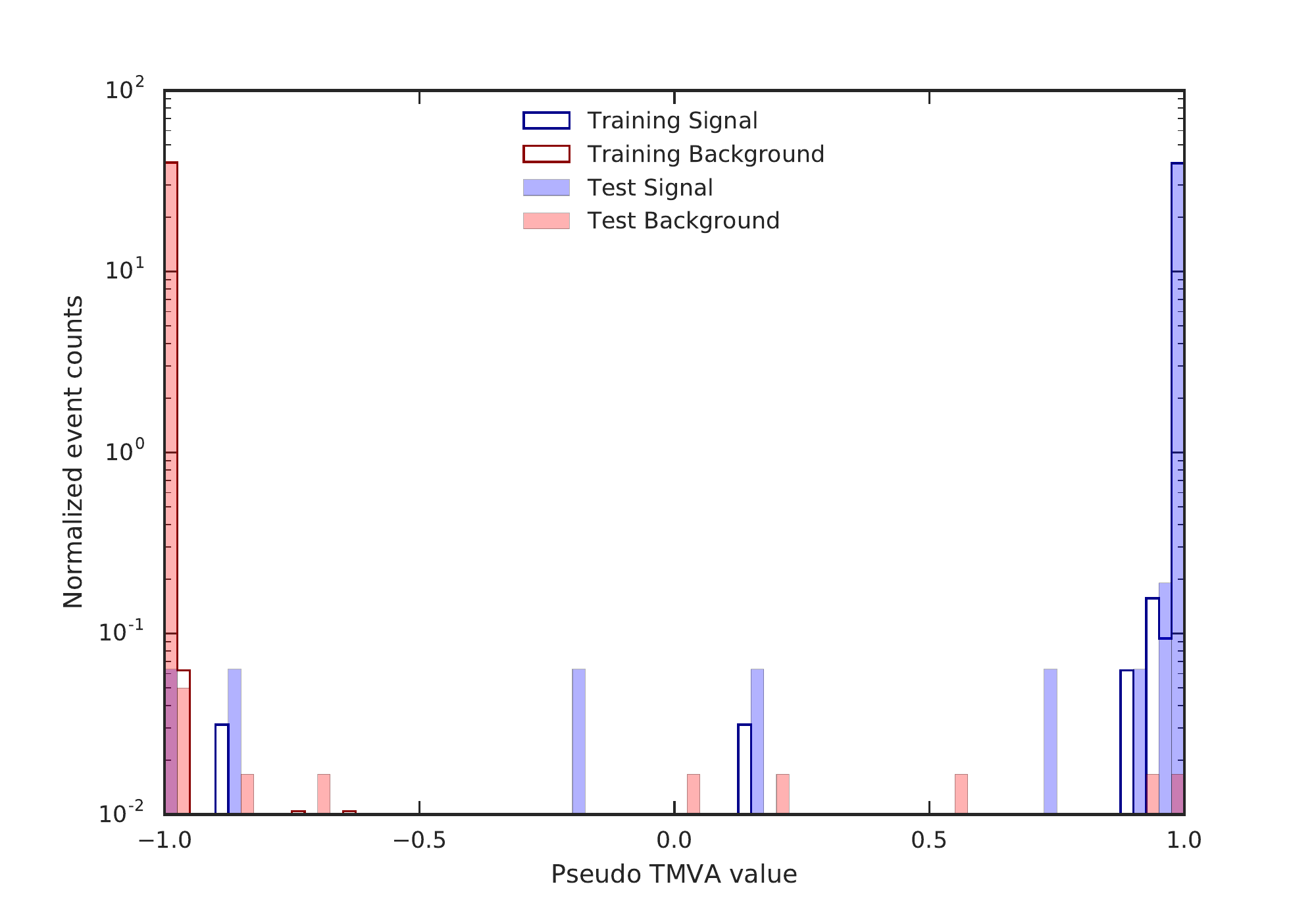} 
 \caption{\small The distribution of the predictions of the training (filled histogram) and the test (step unfilled histogram) events. The signal events are shown in blue, and the background events are shown in red. The X-axis shows a projection of the probability of an event being signal (see text), and the Y-axis is shown in log scale to highlight the few outliers. }
   \label{fig2}
\end{center}
\end{wrapfigure}
%
The CNN model described above was first trained on randomly-selected 70\% of the training data, and cross-validated on the remaining 30\% of the training data. The binary cross-entropy is used as the cost function, which is optimized using stochastic gradient descent with a mini-batch size of 128 and a momentum of 0.9. The learning rate is initially set at 0.01, and decays by $10^{-6}$ after each iteration. We set an early stop criterion so that the training stops if the cost function stops improving for 10 epochs. The training process stopped after 38 epochs, which took roughly 5 minutes running on an NVIDIA\textregistered Tesla\textsuperscript{TM} C2050 GPU. This trained model achieved a training AUC score of 0.999995, and a validation AUC score of 0.999925. 

To further evaluate the trained CNN model, we passed the independent test dataset through the model, and obtained a test AUC score of 0.999962. 
Fig.\,\ref{fig2} shows the histograms of the predictions from the trained model. We linear transform the probability $P$ of an event being a signal muon event to $(2P-1)$, to be consistent with the output convention in the \textit{ROOT TMVA} package \citep{Brun97} used in previous work on gamma-hadron separation. 
The good agreement between the training and the test histograms 
indicates that overfitting is minimal. 

\begin{figure}[h]
\begin{subfigure}{.5\textwidth}
\begin{center}
  \includegraphics[width=1\textwidth]{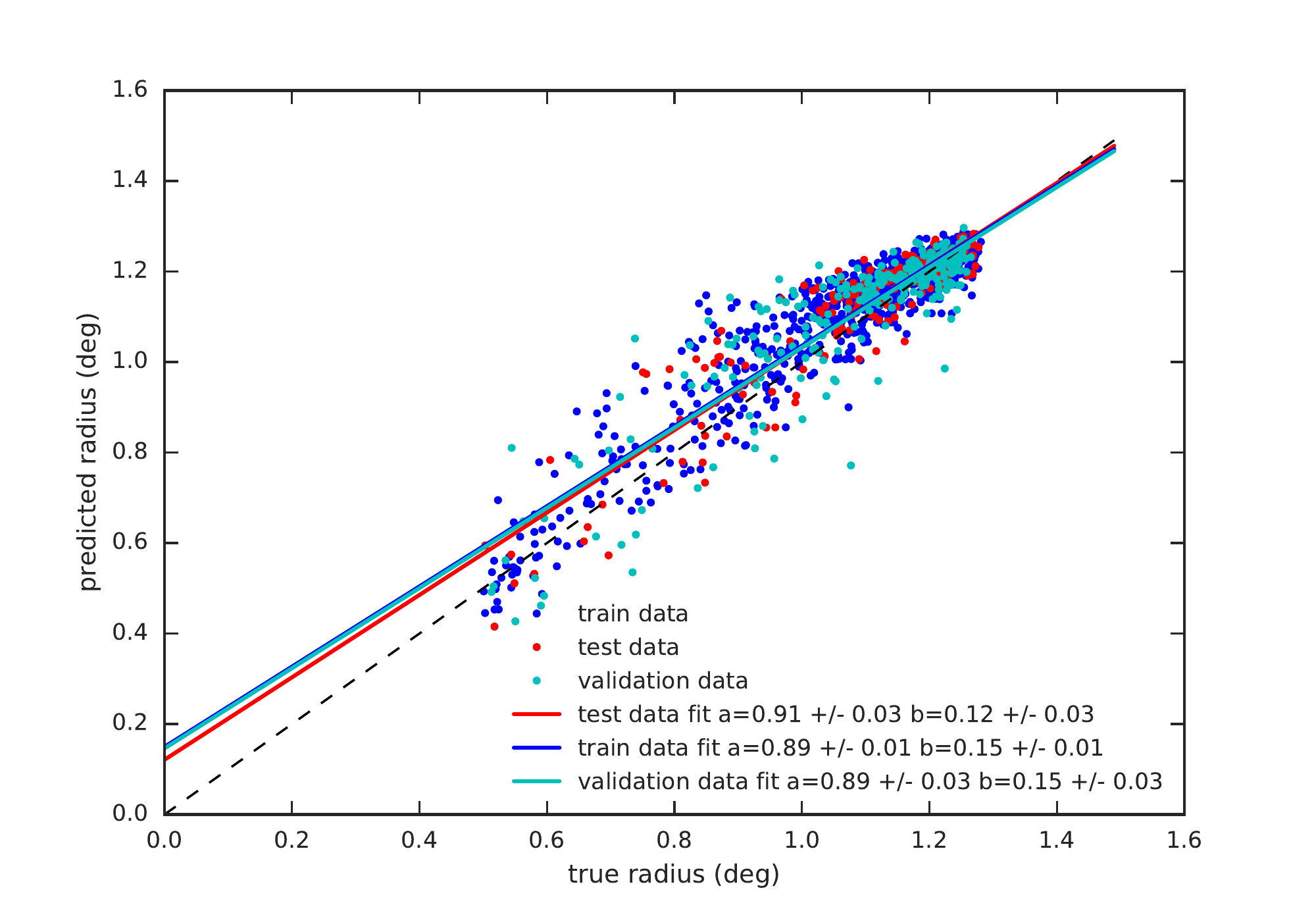} 
 \caption{\small The input radius vs. the predicted radius.}
   \label{figRadPred}
\end{center}
\end{subfigure}%
\begin{subfigure}{.5\textwidth}
\begin{center}
 \includegraphics[width=1\textwidth]{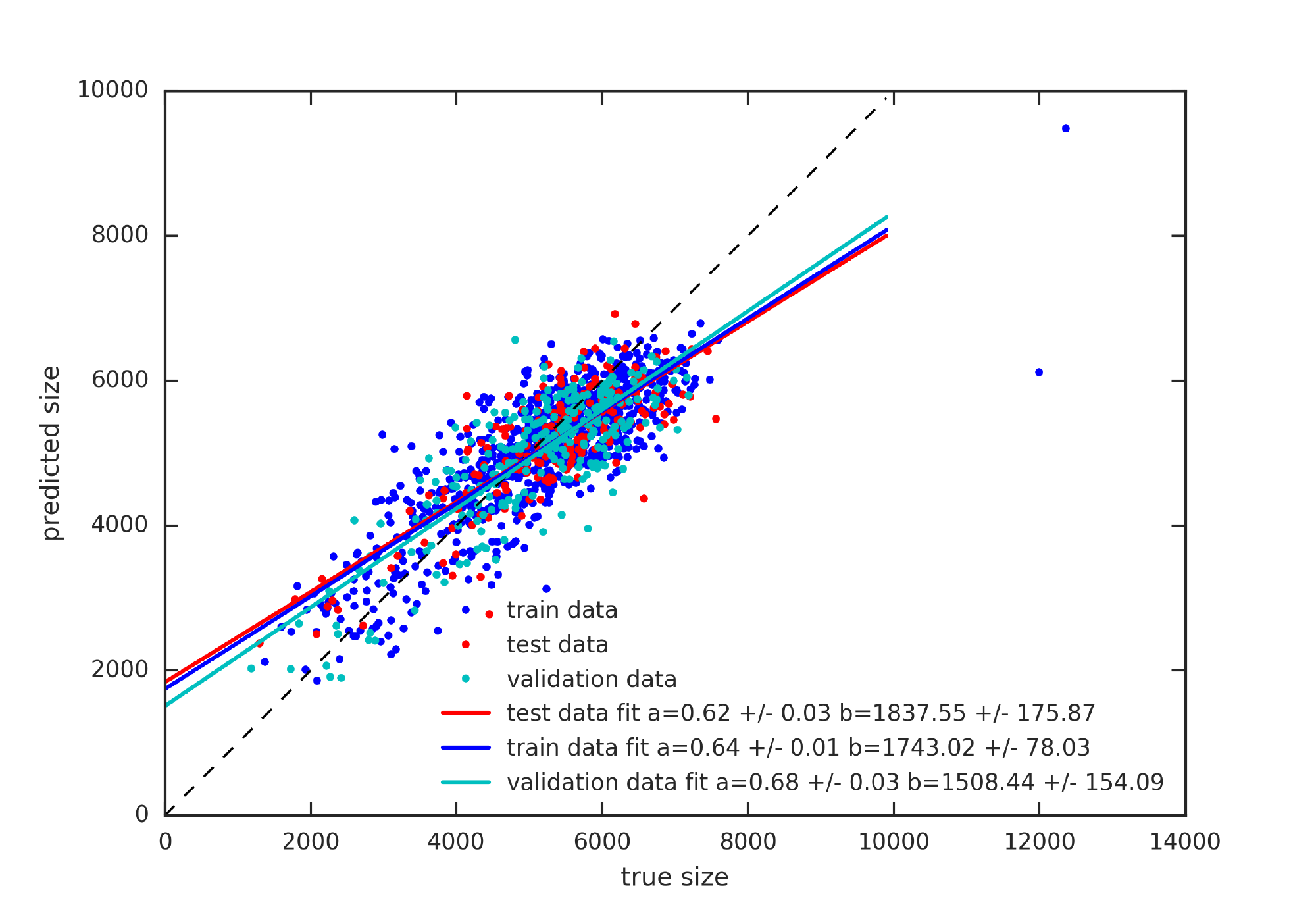} 
 \caption{\small The input size vs. the predicted size.}
   \label{figSizePred}
\end{center}
\end{subfigure}%
 \caption{\small The input values vs the prediction values of the muon-ring radius and brightness. Each dot corresponds to a muon event. The training events are shown in blue, the validation events are shown in cyan, and the test events are shown in red. The solid lines are the best-fit linear models, the slope $a$ and the intercept $b$ of which are shown in the legend. }
\end{figure}

 \vspace*{-0.5 cm}
\section{The CNN regression model}
To build a CNN regression model for predicting continuous variables, we first train a CNN classifier following the procedure described in the previous section. 
Then we extract only the convolutional layers from the trained CNN classifier model. 
The VERITAS data used for the regression training and testing are from two 30-minute observations different from the classifier training/test data. 
The new training data are processed through the trained convolutional layers, and the output of the convolutional layers is used as the input to a new fully-connected regressor model. 
We performed a crude grid search and construct the regressor as two fully-connected layers with 2048 and 64 neurons, respectively. We apply dropout at 30\% ratio after each of the two layers. 
Rectified linear unit is used as the activation function of both layers, as well as the output activation. 

Two regressor models on the radius of the muon rings and the brightness (or size as measured in digital counts) of the muon images, respectively, are built and trained. 
The training process is similar to the classifier model. 
The mean squared error (MSE) is used as the cost function, and the \textit{Adam} optimizer \citep{Kingma14} with an initial learning rate of $10^{-3}$ and exponential-decay parameters $\beta_1=0.9$ and $\beta_2=0.999$. The weights are updated in mini batches of size 128. We set an early stop criterion so that the training stops if the cost function stops improving for 50 epochs. 
The model setup and training for the brightness regressor is similar. 

The scatter plots of the ``true'' radius and size calculated by \textit{VEGAS} versus the predicted radius and size using the two trained regressor models are shown as Fig.\,\ref{figRadPred} and Fig.\,\ref{figSizePred}, respectively. A general linear correlation is apparent, although with a noticeable amount of variance. The model tends to over-predict smaller values and under-predict larger values, as indicated by the best-fit slope $<1$. The regressor for the brightness performs worse compared to that for the radius, likely due to the lack of image cleaning. The relatively large variance and bias in the model need to be addressed. 

 \vspace*{-0.3 cm}
\section{Implications}
Using the simple case of muon-ring images as a pathfinder for gamma/hadron separations, we demonstrated that it is possible to use the raw images of an IACT as a direct input into a state-of-the-art CNN model and obtain very accurate classification results. It has the potential to exploit subtle features in the Cherenkov shower images, while being resistant to noise even without image cleaning. Moreover, with the ability of regression, there is a potential to use CNN models for the direction and energy reconstructions of the primary gamma rays, and produce high-level scientific results. It is worth noting that the prediction is fast once the model is trained, therefore such reconstructions can be realtime. 

For future work, we plan to investigate the effect of adding different levels of image cleaning in the preprocessing, improve the regressor models so that they can be used for the calibration of throughput efficiency of VERITAS, and build classifier models for gamma-hadron separation, the biggest challenge of which is to obtain reliable labels of gamma-ray events. Crowdsourcing like \href{www.zooniverse.org}{\textit{Zooniverse}} \cite[e.g.][]{Simmons16} provides one viable solution. 

 \vspace*{-0.3 cm}

\begin{acknowledgement}
{\contribsize
This research is supported by grants from the U.S. Department of Energy Office of Science, the U.S. National Science Foundation and the Smithsonian Institution, and by NSERC in Canada. We acknowledge the excellent work of the technical support staff at the Fred Lawrence Whipple Observatory and at the collaborating institutions in the construction and operation of the instrument. \\
\indent The VERITAS Collaboration is grateful to Trevor Weekes for his seminal contributions and leadership in the field of VHE gamma-ray astrophysics, which made this study possible.
}
\end{acknowledgement}

\end{document}